\documentclass[apj]{emulateapj}
\usepackage{apjfonts}
\usepackage{epsfig}
\usepackage{subfigure}
\usepackage{amsmath}
\usepackage{amssymb}
\usepackage{amsfonts}
\usepackage{natbib}          
\lefthead{Paladini et al.}
\righthead{}

\begin{document}

\title{The {\em{Planck}} submillimeter properties of Galactic high-mass star forming regions: 
dust temperatures, luminosities, masses and Star Formation Efficiency}

\author{
R. Paladini\altaffilmark{1,*},
J. C. Mottram\altaffilmark{2},
M. Veneziani\altaffilmark{3},
A. Traficante\altaffilmark{4},
E. Schisano\altaffilmark{4},
G. Giardino\altaffilmark{5},
E. Falgarone\altaffilmark{6},
J. S. Urquhart\altaffilmark{7}
D. L. Harrison\altaffilmark{8,}\altaffilmark{9},
G. Joncas\altaffilmark{10},
G. Umana\altaffilmark{11},
S. Molinari\altaffilmark{4},
}

\altaffiltext{*}{Corresponding author: paladini@ipac.caltech.edu}
\altaffiltext{1}{
Infrared Processing Center, California Institute of Technology,
              1200, E. California Blvd., Pasadena, CA 91125,  USA
}
\altaffiltext{2}{Max Planck Institute for Astronomy, Konigstuhl 17, 69117 Heidelberg, Germany
}

\altaffiltext{3}{Science $\&$ Technology corporation,  Olof Palmestraat 14, 2616 LR Delft, The Netherlands 
}

\altaffiltext{4}{IAPS - INAF, Via Fosso Del Cavaliere 100, I-00133, Roma, Italy
}

\altaffiltext{5}{ESA, Science Operations Department, ESTEC, 2200AG Noordwijk, 
The Netherlands}

\altaffiltext{6}{LERMA/LRA, Observatoire de Paris, PSL Research University, CNRS, Sorbonne Universites,
UPMC, Universite Paris 06, Ecole Normale Superiore, 75005 Paris, France
}

\altaffiltext{7}{Centre for Astrophysics and Planetary Science, University of Kent, Canterbury, CT2 7NH, UK
}

\altaffiltext{8}{Institute of Astronomy, Madingley Road, Cambridge, CB3 0HA, UK}

\altaffiltext{9}{Kavli Institute of Cosmology Cambridge, Madingley Road, Cambridge,
CB3 OHA, UK}

\altaffiltext{10}{Universite' Laval, Pavillon Alexandre-Vachon 1045, Avenue de la Medecine, Canada
}

\altaffiltext{11}{INAF - Osservatorio Astrofisico di Catania,
         Via S. Sofia 78, 95123 Catania, Italy
}

\begin{abstract}

Massive star formation occurs in the interior of giant molecular clouds (GMC) and proceeds through many stages. 
In this work, we focus on massive young stellar objects (MYSOs) and Ultra-Compact HII regions (UCH\,{\sc ii}), where the former 
are enshrouded in dense envelopes of dust and gas, which the latter have begun dispersing.  
By selecting a complete sample of MYSOs and UCH\,{\sc ii} from the Red MSX Source (RMS) survey data base, we combine {\em{Planck}} and 
IRAS data and build their Spectral Energy Distributions (SEDs). With these, we estimate the physical properties 
(dust temperatures, mass, luminosity) of the sample. Because the RMS database provides unique solar distances, it also allows 
investigating the {\em{instantaneous}} Star Formation Efficiency (SFE) as a function of Galactocentric radius. 
We find that the SFE increase between 2 and 4.5 kpc, where it reaches a peak, likely in correspondence of the accumulation of 
molecular material at the end of the Galactic bar. It then stays approximately constant up to 9 kpc, after which it linearly 
declines, in agreement with predictions from extragalactic studies. This behavior suggests the presence of a significant amount of undetected molecular gas at R$_G$ $>$ 8 kpc. 
Finally we present diagnostic colors that can be used to identify sites of massive star formation. 

\end{abstract}


\section{Introduction}

\begin{figure*}{H}
\centering
\includegraphics[width = 19cm, height = 13.5cm,angle=0]{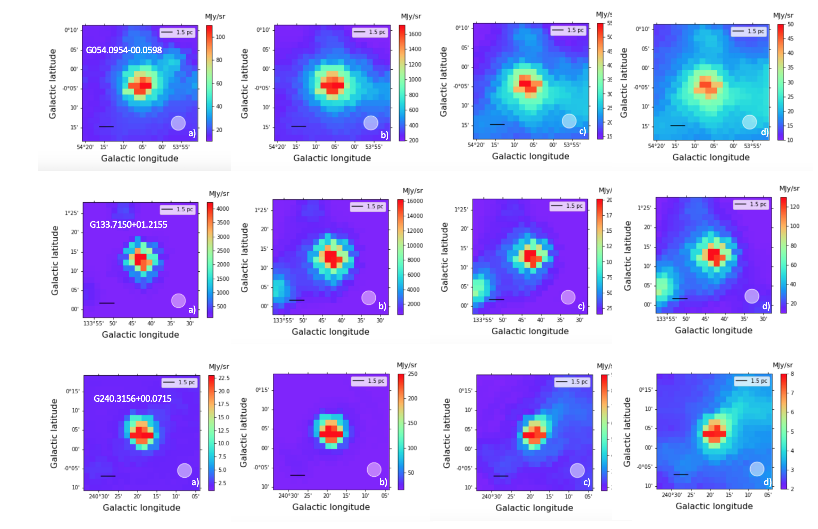}
\caption{Examples of MYSOs and UCH\,{\sc ii} regions identified in the {\em{Planck}} data at the RMS position. From left to right: a) IRIS 25 $\mu$m; b)
IRIS 100 $\mu$m; c) {\em{Planck}} HFI 857 GHz (350 $\mu$m); d) {\em{Planck}} HFI 353 GHz (850 $\mu$m). All the data are convolved to the common 
{\em{Planck}} HFI 353 GHz angular resolution (4.86$^{\prime}$).}
\end{figure*}

The initial stages of a massive star can be traced back to a giant molecular cloud (GMC), i.e. a cloud with a mass 
ranging from 10$^{5}$ M$_{\odot}$ to 10$^{6}$ M$_{\odot}$ (e.g Dame et al. 2001; Miville-Deschenes et al. 2017) and a linear size up 
to hundreds of pc. Within GMCs, very dense molecular cores (n $>$ 10$^{5}$ cm-3, e.g. Giannetti et al. 2013) collapse, and evolve first into massive young stellar objects (MYSOs), and later, when  
the OB star begins to ionize the surrounding material, into Ultra-Compact H\,{\sc ii} regions (UCH\,{\sc ii}). 

In this framework, characterizing clumps hosting massive star formation, an intermediate stage between GMCs and cores, is of primary importance: massive star formation is 
known to take place in cold (T$_d <$ 25 K), massive (M $>$ 100 M$_{\odot}$), luminous (L $>$ 10$^{3}$ L$_{\odot}$) environments, so constraining dust temperatures, luminosities 
and masses of the clumps allows one to assess whether the clump under investigation is able to effectively form massive stars. In general, determining accurate masses and 
luminosities for the clumps is also crucial to assess their evolutionary stage on a mass-luminosity plot (e.g. \cite{molinari}). 

In this work we are going to use the combined {\em{Planck}} High Frequency Instrument (HFI) (Tauber et al. 2010, 
Planck Collaboration 2011a, Lamarre et al. 2010; Planck HFI Core Team 2011a) and IRAS (Neugebauer et al. 1984) data 
to investigate a complete sample of clumps harboring MYSOs and UCH\,{\sc ii} regions. This study has two primary objectives: 1) to derive the properties (dust temperature, 
luminosity, mass, surface density) of these clumps and therefore fully characterize this evolutionary stage; 2) to make use of the estimated luminosities and masses, to 
compute an {\em{instantaneous}} Star Formation Efficiency (SFE) for the first 0.1 Myr (the approximate age of UCH\,{\sc ii} regions). For 2), we aim at 
addressing the question: At this fixed evolutionary stage, is there variation in the amount of gas relative to the star-formation (modulo the spread caused 
by the variation in age ?) And, what could cause variations in the efficiency that gas is converted to stars: galactic dynamics (sheer, bars), star formation 
mechanisms (cloud-cloud collisions), or hidden gas (e.g. undetected molecular material) ?

The combination of the spectral bands from {\em{Planck}} and IRAS is extremely effective in probing 
the Spectral Energy Distribution (SED) of sources such as MYSOs and UCH\,{\sc ii}. The coldest dust component present in these objects typically thermalizes at temperatures of the order of 15 - 20 K, and 
this makes the 850 $\mu$m {\em{Planck}} band a very accurate tool in estimating the total mass. The accuracy of the 
{\em{Planck}} mass measurements is also due to the fact that this experiment operated from space, which means that no filtering for atmospheric noise is applied at any stage of data processing, 
hence  emission on all angular scales is preserved. This is not the case for ground experiments such as APEX/LABOCA, JCMT/SCUBA and CSO/Bolocam, where considerable flux loss occurs, with 
effects on mass determination. An attempt to combine {\em{Planck}} and APEX/LABOCA data was carried out for the APEX Telescope Large Area Survey of the Galaxy (ATLASGAL, 
Schuller et al. 2009), a large and sensitive sub-millimeter survey of the inner Galactic plane. This work (Csengeri et al. 2016) though was limited to the investigation of the large-scale 
structure of cold dust.

Noteworthy, {\em{Planck}} and IRAS both mapped the whole sky, including both the inner and outer Galactic Plane. 
The outer Galactic disk (R$_G$ $>$ 8.5 kpc) has characteristics that sets it apart from the inner Galaxy. For instance, for R$_G$ $>$ 13 kpc, both 
the H$_{2}$ surface density (Scoville $\&$ Sanders 1987; Digel et al. 1996; Heyer et al. 1998) and the stellar disk (Robin et al. 1992; Ruphy et al. 1996) appear 
to exhibit a sharp decline. In addition, the metallicity, at these large Galactocentric distances, is only half solar ($\sim$ 0.5 Z$_{sol}$, Yong et al. 2005). 
These observational facts suggest a dramatic change in star formation activity with respect to the inner Galaxy.

We emphasize that the {\em{Planck}} and IRAS beams (of the order of 5 arcmins) are sensitive to structures of $\sim$ 1.5 pc (at a distance of 1.5 kpc), which is 
the typical size of Galactic clumps. Although clumps will contain multiple 
cores, we expect MYSOs or UCH\,{\sc ii} regions to be the dominant component within each individual clump.

The paper is organized as follows. Section~2 describes the {\em{Planck}} and the IRAS data sets.  Section~3 provides details on the samples selection and 
on the identification of the sources in the {\em{Planck}}
data. Section~4 discusses the photometric measurements, the SEDs, as well as the estimate
of the clump dust temperatures, luminosities, masses and surface densities. Section~5 presents the Star Formation Efficiency in terms of clump luminosity-to-mass ratio and its variations
across the Galaxy. A summary is provided in the Conclusions (Section~6). 
Finally the Appendix contains an estimate of the colors that can be used as diagnostics to identify regions of massive star formation.

\section{Planck data}

{\em{Planck}}{\footnote{{\em{Planck}} (http://www.esa.int/Planck)
is a project of the European Space Agency (ESA) with instruments provided by two scientific consortia
funded by ESA member states (in particular the lead countries France and Italy), with contributions
from NASA (USA) and telescope reflectors provided by a collaboration between ESA and a scientific
consortium led and funded by Denmark.} (Tauber et al. 2010; Planck Collaboration 2011a) was the third-generation
space mission to measure the anisotropy of the Cosmic Microwave Background (CMB). It observed
the sky in nine frequency bands covering 30 - 857 GHz (i.e., 10,000 $\mu$m - 350 $\mu$m), with high sensitivity and angular resolution
from 31$^{\prime}$ to 5$^{\prime}$. The Low Frequency Instrument (LFI, Mandolesi et al. 2010) covered the 30, 44 and 70 GHz bands with amplifiers cooled
to 20 K. The High Frequency Instrument (HFI, Lamarre et al. 2010; Planck HFI Core Team 2011a) covered 
the 100, 143, 217, 353, 545 and 857 GHz bands with bolometers cooled to 0.1 K.
Polarisation was measured in all but the two highest frequency bands (Leahy et al. 2010; Rosset et al. 2010).
{\em{Planck}}'s sensitivity, angular resolution, and frequency coverage made it a powerful instrument for Galactic and extragalactic astrophysics as well
as cosmology. 

In this paper we use the {\em{Planck}} PR2 full channel, full mission temperature maps
at nominal frequencies 353 (850 $\mu$m), 535 (560 $\mu$m) and 857 GHz (350 $\mu$m). 
We downloaded the maps from the IRSA archive{\footnote{https://irsa.ipac.caltech.edu/data/Planck/release$\_$2/}}. 
The 353 GHz map are in T$_{CMB}$ units, while the 535 and 857 GHz maps are in MJy/sr. To perform the conversion from 
CMB thermodynamic units to Rayleigh-Jeans brightness temperature units we make use of the conversion factor given 
in Table~3 of Planck Collaboration.X. (2016). 

We adopt the effective beam sizes provided in the Explanatory Supplement{\footnote{https://wiki.cosmos.esa.int/planck$-$legacy$-$archive/index.php/Effective$\_$Beams}}. 
These are: 4.86$^{\prime}$, 4.84$^{\prime}$ and 4.63$^{\prime}$ at, respectively, 353, 545 and 857 GHz. 
 
\subsection{Ancillary data}

In order to construct SEDs of the clumps, we complement the {\em{Planck}}
data with IRAS (Neugebauer et al. 1984) data at 100, 60 and 25$\mu$m. In particular, we use the IRIS (Improved
Reprocessing of the IRAS Survey, Miville-Deschenes $\&$ Lagasche 2005) maps, which benefit from the COBE-DIRBE (Hauser et al. 1998)
calibration and zero point, as well from a better zodiacal light subtraction and  destriping.
We do not include the IRIS 12$\mu$m data since this band is significantly contributed to by emission
from Polycyclic Aromatic Hydrocarbons (PAHs), whose modeling and characterization is beyond the scope
of this paper. The Full-Width-Half-Maximum (FWHM) of the IRIS beams are, respectively, 3.8', 4.0' and 4.3', at 25, 60 and 100 $\mu$m.\\ 

Both the {\em{Planck}} and IRIS data are reprojected into the HEALPIX (Hierarchical Equal Area isoLatitude Pixelization, Gorski et al., 2005) format
with Nside = 2048. All maps are smoothed to the {\em{Planck}} 353 GHz angular resolution (4.86$^{\prime}$).

\section{Sample selection}
\label{sec:obs}

Our reference data base for the source selection is the RMS{\footnote{http://www.ast.leeds.ac.uk/RMS/}} survey
(Lumsden et al. 2013), i.e. a campaign of follow-up observations designed to identify possible contaminants (e.g. evolved stars and planetary nebulae, PNe)
of a sample of $\sim$ 2000 candidate MYSOs assembled by Lumsden et al. (2002) using MSX color
selection criteria. Examples of the follow-up observations which have been
undertaken by the RMS team are high-resolution (1'') radio continuum observations
(Urquhart et al. 2007a, 2009b) which have allowed the descrimination of UCH\,{\sc ii} regions from
PNe, and near-IR spectroscopic measurements which have led to the
identification of evolved stars (Clarke et al. 2006). This effort has made it possible to
single out some 1500 (1420) MYSOs and UCH\,{\sc ii} regions  with 
uniquely constrained distances (Urquhart et al. 2013) . Radial velocities have been obtained from $^{13}$CO J = 1 - 0 and J = 2 -1 observations
(Urquhart et al. 2007b, 2008b), and these have been coupled with the Reid et al. (2009) rotation curve to derive
kinematic distances. 

See Figure~1 for examples of MYSOs and UCH\,{\sc ii} regions from the RMS database in the IRAS/IRIS and {\em{Planck}} data.

Mottram et al. (2011) investigate the completeness of the RMS sample. To this end,
they compute the volume of the Galaxy probed by the RMS survey at each luminosity $L$ and assume
that the RMS MYSOs and HII regions are distributed in a layer with a hole in the center, according to the model by Robin et al. (2003) for the thin disk Galactic stellar population.
Following this method, they obtain that the survey is 100$\%$ complete for luminosities greater than $\sim$ 1$\times$10$^{4}$ L$_{\odot}$. This selection leads to 731 sources.

The longitude and latitude distributions of the complete sample are shown in Figure~2. From the longitude
distribution, it is evident that there is a higher source concentration towards the intersection of the line of sight with the Sagittarius and Scutum-Crux arms 
(first Galactic quadrant, {\em{l}} $\sim$ 30$^{\circ}$) and with the Norma and Scutum-Crux arms (fourth Galactic quadrant, {\em{l}} $\sim$ 330$^{\circ}$). 
Out of 731 sources, 209 are MYSOs candidate, 509 are UCH\,{\sc ii} regions, 13 are thought to be transition objects, i.e. sources that are older than MYSOs but not old enought to be 
classidied as UCH\,{\sc ii} regions. 
In addition, 151 ($\sim$ 20 $\%$) are located at Galactocentric distances greater than 8.5 kpc, and 108 are part of complexes of multiple sources.

\begin{figure*}
\centering
\includegraphics[width=8cm,height=6cm]{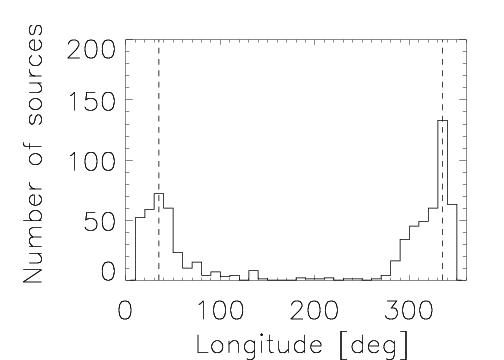}
\includegraphics[width=8cm,height=6cm]{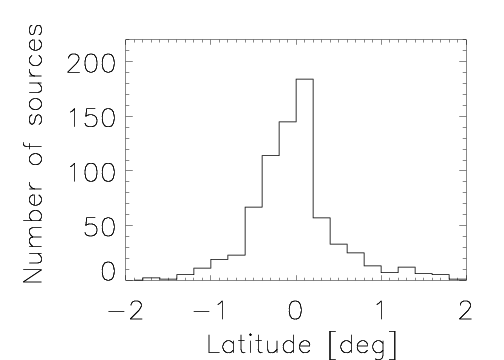}\\
\caption{Longitude (left panel) and latitude (right panel) distributions of the sample of candidate MYSOs and UCH\,{\sc ii} regions above the 10$^{4}$ L$_{\odot}$ completeness limit. The dashed lines 
on the longitude plot denote, respectively, the Sagittarius and Scutum-Crux arms (at {\em{l}} $\sim$ 30$^{\circ}$) and the Norma and Scutum-Crux arms (at {\em{l}} $\sim$ 330$^{\circ}$).}
\end{figure*}

For each source in the RMS data, we create a 30$^{\prime}\times$30$^{\prime}$ image centered on the source at each
{\em{Planck}} frequency (353, 545, 857 GHz). Each image is visually inspected to assure that the emission in the considered region is not dominated
by background emission.  From this visual inspection we notice that strong
sources of emission centered on the location of the RMS source are, with a few exceptions ($\sim$ 5$\%$), present in the {\em{Planck}} HFI bands.  

To estimate the angular size of these structures, we follow two complementary approaches. First we assume that all the sources are unresolved with respect 
to the {\em{Planck}} 353 GHz convolution beam (4.86$^\prime$), second we perform a 2-dimensional gaussian fits and, for each clump, we derive
a measure of the major and minor axis at 353 GHz (850 $\mu$m). In this case, not all the sources turn out to be resolved. Out of 731, only for 238 (32.5 $\%$) we can 
determine the major and minor axis, with an average size of 5.6' $\pm$ 1.5'. 
In the following sections, we will only show the results for the unresolved case, while we will annotate in parenthesis the results for the fitted size case.

\section{SEDs} 

To compute the flux in each band, we use the HEALpix aperture photometry code developed for Planck Collaboration (2011c). 
As input aperture radius, we use half the convolution beam FWHM (4.86') if the sources are unresolved, otherwise 
we use half of the estimated FWHM from the 2-d gaussian fit. In case the sources are quasi-spherically symmetric, i.e. their aspect ratio is $\sim$ 1, the major and minor axis
are averaged together and half this average is used as the source aperture, {\em{aper$_{s}$}}. Alternatively, half of the major axis is used as the aperture. 

After converting the maps in units of Jy/pixel, the pixels within an aperture equal to {\em{aper$_{s}$}} are summed together. An estimate of the background
is subtracted using a median estimator of the pixels within radii [{\em{aper$_{s}$}}, 2$\times${\em{aper$_{s}$}}]. Uncertainties are obtained by summing in quadrature the r.m.s of the values
in the background annulus to the absolute calibration uncertainties for each map. 

For the sources in double (90) or triple systems (18), in order to avoid double counting, we have subdivided the total flux in the Planck beam by two or three, depending 
on the case. 

Since the pioneering works of Chini et al. (1986a, 1986b, 1986c, 1987),
it is known that, in order to explain observations of HII regions above and below $\sim$ 100$\mu$m,
one has to invoke the existence of a 2-temperature component dust distribution: a warm, low density population
of dust grains situated in the proximity of the central source, and a colder dust population in the periphery of the cloud.  The preliminary finding by Chini et al. was subsequently confirmed
by data at increasingly higher spatial resolution and larger spectral coverage (e.g. Povich et al. 2007), as well as by sophisticated radiative transfer modeling (e.g. Zhang $\&$ Tan 2011). 
The SEDs obtained from the combination of the
{\em{Planck}} and IRIS data show a behaviour similar to the one just described (Figure~3, blue points), as the IRIS 60$\mu$m and 25$\mu$m data points cannot be represented by
the same grey-body as the measurements at longer wavelengths. Taking this fact into account, we fit our SEDs with the functional form:

\begin{equation}
S_{\lambda} = A_{1} \left(\frac{\lambda}{\lambda_{0}}\right)^{-\beta} B_{\lambda}(T_{c}) + A_{2} \left(\frac{\lambda}{\lambda_{0}}\right)^{-\beta} B_{\lambda}(T_{w})
\end{equation}

where T$_{c}$ and T$_{w}$ are, respectively, the temperatures of the cold and warm components, and $\lambda_{0}$ is set to 100 $\mu$m.

To estimate the parameters A$_{1}$, A$_{2}$, T$_{c}$ and T$_{w}$, we use a  $\chi$$^{2}$ goodness-of-fit method. For the spectral emissivity index, $\beta$, we test three 
different values: 1.8, 2.0 and 2.2. The best $\chi$$^{2}$ are consistely given by fits performed with $\beta$ = 1.8 (see Table~1), therefore this is the value that we adopt 
for both grey-bodies in the final runs. 

Note that we do not perform the photometric SED fitting by using a more sophisticated modeling (e.g. Robitaille et al. 2006, 2007) for several reasons: at the angular resolution we are working, 
we are not sensitive to the parameters probed by these models (e.g. disk inner/outer radius, cavity opening angle, etc.); moreover we are probing the extent of clumps, while those frameworks  
were designed for describing the behavior of individual cores; finally UCH\,{\sc ii} represent a far too advanced evolutionary stage not accounted for by those models. 

\begin{table}
\begin{center}
\caption{Spectral emissivity index, $\beta$, and corresponding average SED fitting $\chi^{2}$.}
\begin{tabular}{cc}
\hline
\hline
   $\beta$   & {$\chi^{2}$} \\
    1.8      &    7.3         \\
    2.0      &   15.9          \\
    2.2      &   31.2          \\
\hline
\hline
\end{tabular}
\end{center}
\end{table}

\begin{figure*}
\centering
\includegraphics[width=5cm,height=5cm]{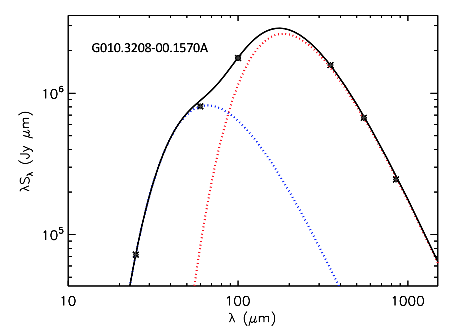}
\includegraphics[width=5cm,height=4.9cm]{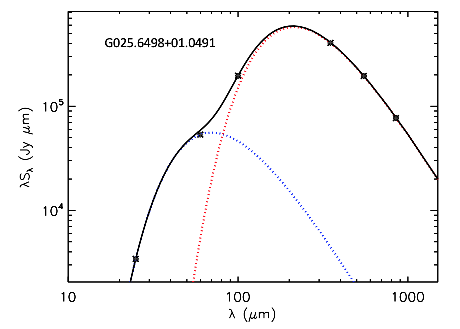}
\includegraphics[width=5cm,height=5cm]{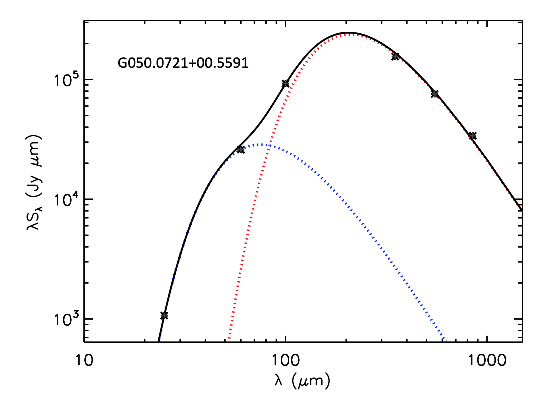}\\
\includegraphics[width=5cm,height=5cm]{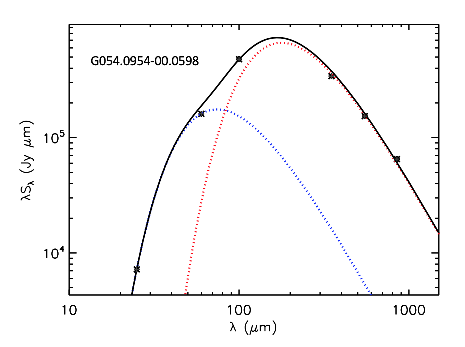}
\includegraphics[width=5cm,height=5cm]{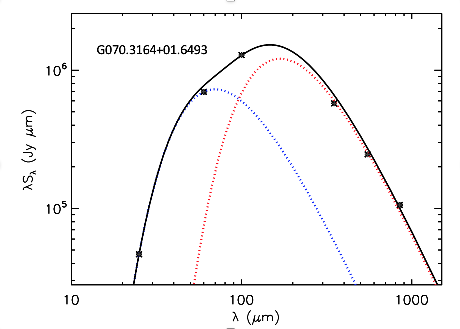}
\includegraphics[width=5cm,height=5cm]{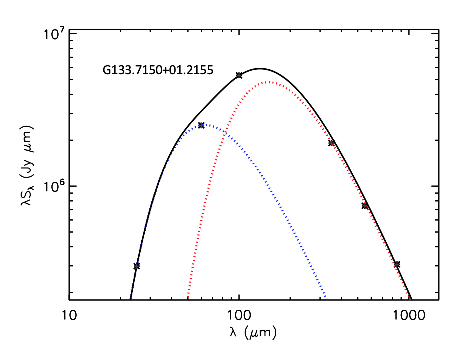}\\
\includegraphics[width=5cm,height=5cm]{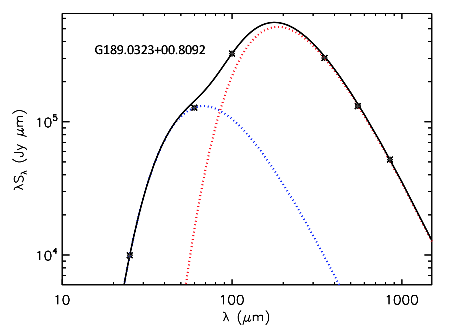}
\includegraphics[width=5cm,height=5cm]{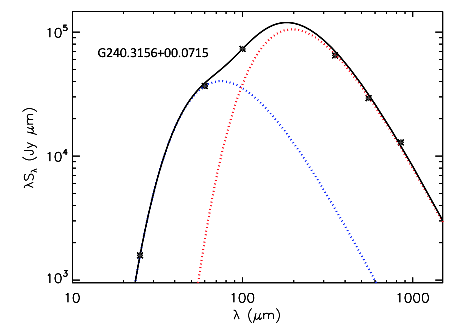}
\includegraphics[width=5cm,height=5cm]{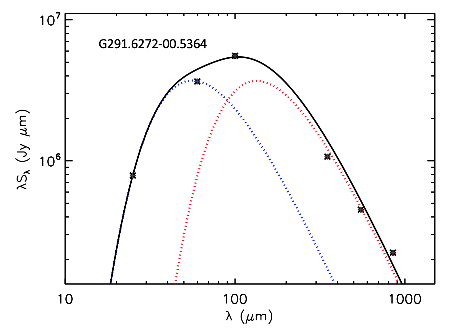}\\
\includegraphics[width=5cm,height=5cm]{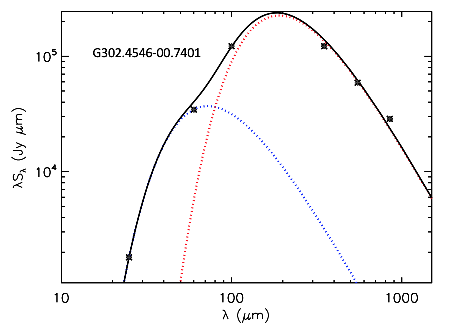}
\includegraphics[width=5cm,height=5cm]{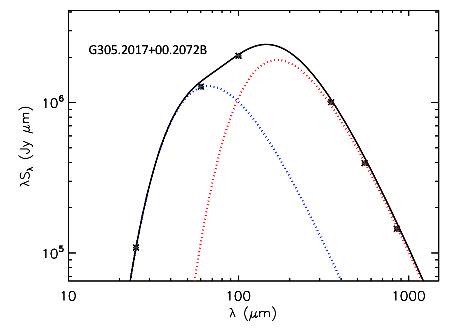}
\includegraphics[width=5cm,height=5cm]{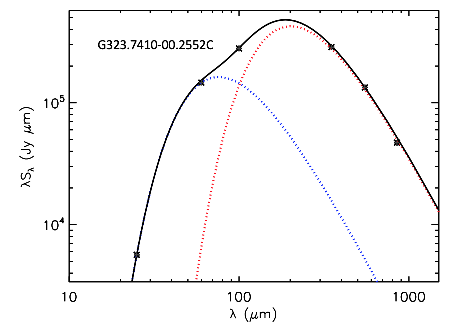}\\
\caption{Examples SED fits to the {\em{Planck}} and IRIS clumps fluxes. The best-fit 2-temperature component model is shown (black solid line). Black diamonds denote the
{\em{Planck}} 350, 500, 850 $\mu$m and IRIS 100, 60 and 25$\mu$m data points. The cold (blue-dashed line) and warm (red-dashed line) temperature components are also shown.}
\end{figure*}

An illustration of the result of the fits is provided in Figure~3 (red solid line). Occasionally, in one (or more) HFI frequency band the best-fit model slightly over-/under-predicts
the measured data points. This is a consequence of the adopted value for $\beta$. We have tested this hypothesis by making trial fits with the spectral emissivity index as a free parameter. Although
in some cases the fit $\chi^{2}$ improves, we prefer to keep $\beta$ fixed and equal to the canonical value of 1.8, as the analysis of variations of $\beta$ and of its degeneracy with respect
to dust temperature (e.g. Schnee et al. 2007; Juvela et al. 2013, 2018) is beyond the scope of this paper.

\subsection{Dust temperatures distribution}

From the fitting procedure, we derive estimates for the cold (T$_{c}$) and warm (T$_{w}$) dust temperatures components, of  
${\overline{T_{c}}}$ = 21.2$\pm$2.9 K 
(21.2$\pm$3.1 K for the fitted size case) and ${\overline{T_{w}}}$ = 54.0$\pm$4.6 K (53.8$\pm$4.4 K for the fitted size case), i.e. comparable to those 
found for more evolved HII regions (e.g. Povich et al., 2007; Paladini et al. 2012). For the uncertainties, rather than the modeling errors, 
we adopt the standard deviations of the sample, given that the former underestimate the true errors. K\"onig et al. (2017) investigate a small sample of MYSOs (36 sources) and UCH\,{\sc ii} regions (25 sources), and 
for these they derive dust temperatures from the modelling of the combined MSX (Egan et al. 2003), WISE (Wright et al. 2010), Hi-GAL and ATLASGAL data. They obtain 
mean temperatures of 28.1$\pm$3.6 K and 31.7$\pm$4.0 K for MYSOs and UCH\,{\sc ii} regions, respectively, which lie in between our average cold and warm components. This 
work was extended to the whole ATLASGAL sample by Urquhart et al. (2018) who find, for MYSOs and HII regions, values of dust temperature between $\sim$ 15 and 40 K.   

We subdivide the sample into 6 Galactocentric bins of equal width (i.e. 2 kpc) and for each bin we compute the average T$_{c}$ and T$_{w}$ (see Table~2).
The distributions of T$_{c}$ and T$_{w}$ vs. R$_{G}$ are plotted in Figure~4 (top panel), and indicate a decreasing temperature for the cold component towards the 
outer Galaxy, in particular for R$_{G}$ $>$ 10 kpc, accompanied by an opposite trend for the warm component.   

An important caveat to keep in mind is that, by binning according to distance from the Galactic center, we assume that the Galaxy is radially symmetric and, in doing so, we ignore the presence of spiral arms. To take this effect into account and
explore even further dust temperature variations across the Galaxy, we split the sample according to longitude. We denote {\em{inner-Galaxy}} sources those located either in the
first (0$^{\circ}$ $< l <$ 90$^{\circ}$) or fourth (270$^{\circ}$ $< l <$ 360$^{\circ}$) Galactic quadrants, and {\em{outer-Galaxy}} sources those either in the second or
third Galactic quadrants (90$^{\circ}$ $< l <$ 270$^{\circ}$). We have 699 candidate MYSOs/UCH\,{\sc ii} regions in the {\em{inner}} sample and 32 in the {\em{outer}} one. 
Then, for each dust temperature component (T$_{c}$ or T$_{w}$), we compare the histogram distribution for the {\em{inner/outer Galaxy}} sub-samples (Figure~4, middle panels). In this case 
the average temperatures are: ${\overline{T_{c,i}}}$ = 21.2$\pm$2.9 K (21.2$\pm$3.1 K for the fitted size case), ${\overline{T_{c,o}}}$ = 20.8$\pm$2.4 K (20.6$\pm$2.7 K); 
${\overline{T_{w,i}}}$ = 53.7$\pm$4.4 K (53.5$\pm$4.3 K for the fitted size case), ${\overline{T_{w,o}}}$ = 59.5$\pm$4.6 K (58.8$\pm$4.6 K for the fitted size case), 
where T$_{c,i}$/T$_{c,o}$ and T$_{w,i}$/T$_{w,o}$ are the cold/warm dust temperatures for the {\em{inner/outer Galaxy}} sub-samples, respectively. 
The histogram distributions highlight a trend similar to what is found in Figure~4 (upper panels). 

\begin{figure*}
\centering
\includegraphics[width=7cm, height=8.5cm, angle=90]{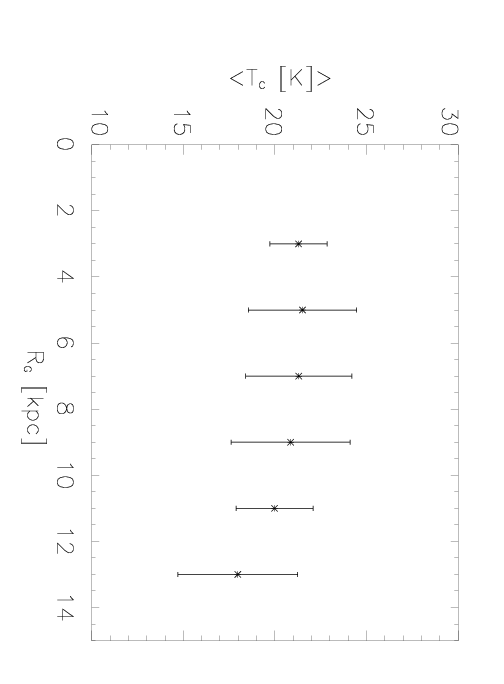}
\includegraphics[width=7cm, height=8.5cm, angle=90]{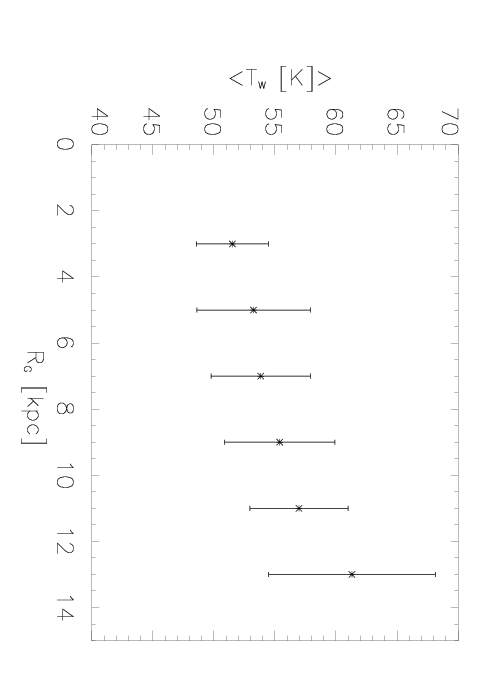}\\
\includegraphics[width=6.5cm, height=8cm, angle=90]{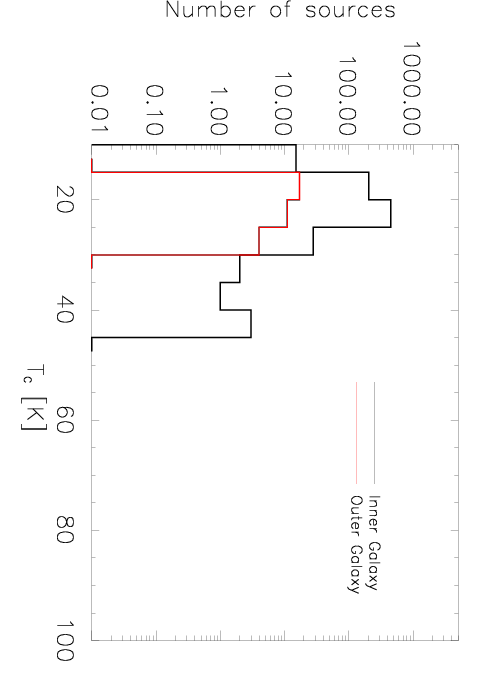}
\includegraphics[width=6.5cm, height=8cm, angle=90]{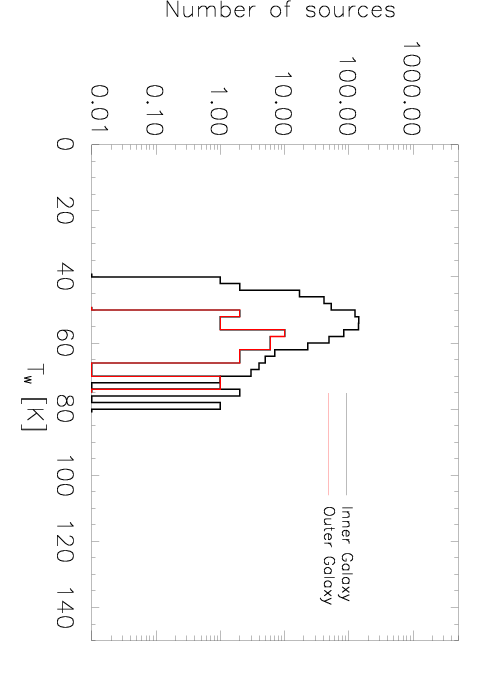}\\
\includegraphics[width=6.5cm, height=8cm, angle=90]{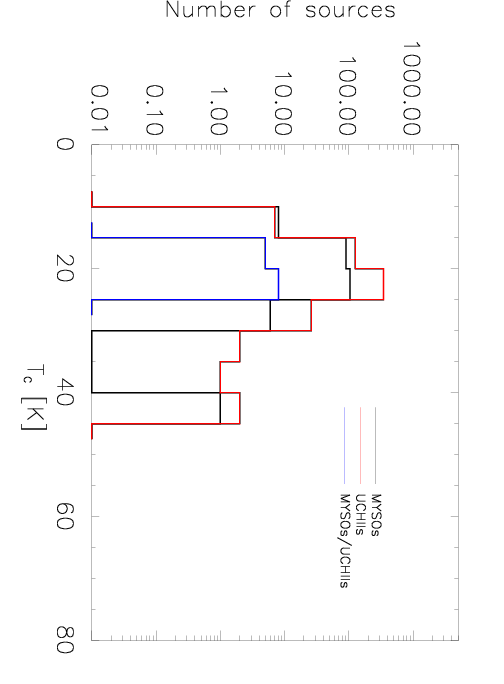}
\includegraphics[width=6.5cm, height=8cm, angle=90]{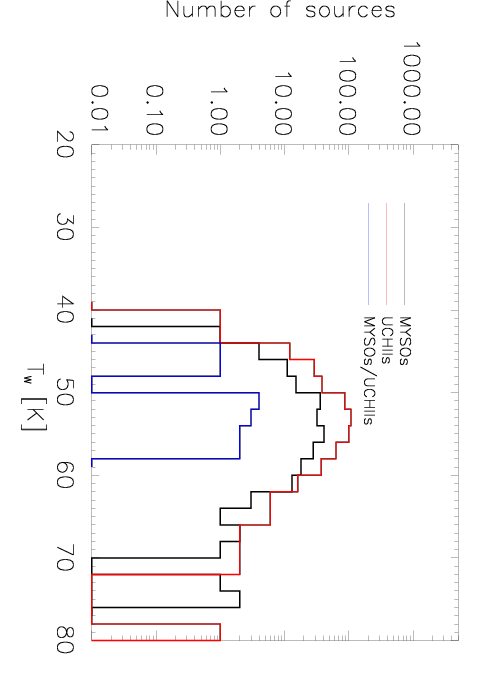}\\
\caption{Top: Average cold (T$_{c}$) and warm (T$_{w}$) dust temperature components as a function of Galactocentric radius. R$_{G}$ values are sampled in 2-kpc bins. Plotted error
bars are the standard deviations in each bin.
Middle: cold and warm dust temperature distributions for sources in the inner (first and fourth) and outer (second and third)
Galactic quadrants. Bottom: cold and warm dust temperature distributions for MYSOs and UCH\,{\sc ii} regions.}
\end{figure*} 

We interpret this result in the  context of the large scale properties of the Galactic plane. A dust temperature gradient is known to characterize the diffuse emission along the Galactic Plane,
with temperatures ranging from 14 - 15 K in the outer Plane, to $\simeq$ 19 K for the inner one (Planck Collaboration 2011m). The temperature
enhancement towards the center of the Galaxy is typically explained as due to the presence of a high concentration of star forming regions, especially in correspondence
of the molecular ring (R$_{G}$ $\sim$ 5 kpc). From our analysis, it appears that the small scale behaviour reflects the situation on larger scales, with clumps harboring
massive stars at large Galactic radii being at lower temperatures with respect to their counterparts closer to the Galactic center. Since T$_{D}$ $\propto$ $X_{ISRF}^{1/(4+\beta)}$ -- with  T$_{D}$ 
denoting the dust temperature,
$X_{ISRF}$ the intensity of the Interstellar Radiation Field (ISRF) and $\beta$ the dust emissivity index --, if we assume a Mathis et al. (1983) radiation field, i.e. scaling with the inverse of 
$R_{G}$, we obtain that
the clumps cold dust component appears to be a local measure of the global radiation field. On the contrary, the clumps warm dust component depends on the stellar radiation field, since this dust
is located in proximity of the young massive stars. Remarkably, an O5 star will heat up dust up to 30K out to a radius of 0.75 pc from the star (Whitney et al. 2005). The trend that we find appears to 
mimic the well-known invertionally proportional increase of electron temperature, T$_{e}$, with galactocentric radius in H$_{II}$ regions (Shaver et al. 1983, Paladini et al. 2004), which is  consequence 
of the metallicity Galactic gradient, and of the fact that metals such as oxygen are coolants. Interestingly, though, we do not see a closer correlation when we split the sample in MYSOs and 
UCH\,{\sc ii} regions. 

A dependance of dust temperature on Galactocentric distance is also reported by Urquhart et al. (2018).  
In their analysis, these authors do not distinguish between cold and warm temperature components, however they find increasing temperatures at larger galactocenctric distances.

Finally, we have analyzed potential differences between the warm/cold dust temperatures of candidate MYSOs and UCH\,{\sc ii} regions (Figure~4, bottom panel).
We do not find any indication that these two populations are characterized by different temperatures. We find: ${\overline{T_{c,MYSO}}}$ = 20.5$\pm$3.0 K (same for the fitted size case),
${\overline{T_{w,MYSO}}}$ = 54.5$\pm$5.0 K (54.3$\pm$4.9 K for the fitted size case); ${\overline{T_{c,UCH_{II}}}}$ = 21.5$\pm$2.8 K (21.5$\pm$3.0 K for the fitted size case), 
${\overline{T_{w,UCH_{II}}}}$ = 53.8$\pm$4.4 K (53.6$\pm$4.2 K for the fitted size case), where T$_{c,MYSO}$/T$_{c,UCH_{II}}$ and
 T$_{w,MYSO}$/T$_{w,UCH_{II}}$ are the cold/warm dust temperatures for the MYSOs/UCH\,{\sc ii} regions.

\begin{table*}
\begin{center}
\caption{Average cold (T$_{c}$) and warm (T$_{w}$) dust temperatures per Galactocentric bin for the two cases of unresolved sources and fitted 
diameters.}
\begin{tabular}{cccccc}
\hline
\hline
   bin center   & n. sources   & T$_{c}$  (unresolved)      &  T$_{w}$ (unresolved)  & T$_{c}$  (fitted sizes) & T$_{w}$ (fitted sizes)  \\
  (kpc)   &              &    (K)                           &  (K)         &                (k)                &    (K)                 \\    
          &              &                                  &              &                    &        \\
    3  &             57       &    21.3$\pm$1.6             &    51.5$\pm$2.9           &   21.2$\pm$1.5                &    51.5$\pm$2.8      \\
    5  &             274      &    21.5$\pm$2.9             &    53.2$\pm$4.6             &  21.6$\pm$3.2                  &  53.1$\pm$4.5     \\
    7  &             224      &    21.3$\pm$2.9             &    53.8$\pm$4.1              &  21.3$\pm$3.0               &    53.7$\pm$3.9       \\
    9   &            126      &    20.8$\pm$3.2             &    55.4$\pm$4.5                &  20.7$\pm$3.3             &    55.1$\pm$4.2         \\
   11  &             39       &    19.9$\pm$2.1             &    56.9$\pm$4.0                &   19.9$\pm$2.4            &    56.5$\pm$4.1        \\
   $>$ 13  &         11       &    17.9$\pm$3.2             &    61.3$\pm$6.8                &   18.1$\pm$3.4            &    60.7$\pm$7.3\\
\hline
\hline
\end{tabular}
\end{center}
\end{table*}

\subsection{Luminosities, masses and surface densities}

Having fitted the source SEDs, we can compute their luminosities. To this end, we use the relation:

\begin{equation}
L_{IR} = 4 \pi D^{2} \int_{\lambda_{min}}^{\lambda_{max}} { S_{\lambda} d\lambda}
\end{equation}

where $S_{\lambda}$ is the best-fit 2-temperature component model discussed in Section~4, and D is the solar distance from the RMS
data base. The integration is performed between 25$\mu$m and 850$\mu$m, i.e. the shortest/longest IRIS/Planck wavelength considered
in our analysis. The result is shown in Figure~5. We obtain ${\overline{L_{IR}}}$ = $8.3\times10^{5}\substack{+1.4\times10^{6} \\ -6.5\times10^{4}}$ L$_{\odot}$ 
(${\overline{L}}$ = $1.1\times10^{6}\substack{+1.7\times10^{6} \\ -6.8\times10^{4}}$  L$_{\odot}$ for the fitted-sizes case). We find that sources in the inner Galaxy are 
on average more luminous than in the outer Galaxy, with ${\overline{L_{IR, inner}}}$= $8.9\times10^{5}\substack{+1.5\times10^{6} \\ -6.9\times10^{4}}$ L$_{\odot}$ 
(${\overline{L_{IR, inner}}}$ = $1.2\times10^{6}\substack{+1.8\times10^{6} \\ -7.6\times10^{4}}$ L$_{\odot}$ for the fitted-sizes case) 
and ${\overline{L_{IR, outer}}}$ = $3.0\times10^{5}\substack{+6\times10^{5} \\ -2.8\times10^{4}}$ L$_{\odot}$ (${\overline{L_{IR, outer}}}$ = $3.6\times10^{5}\substack{+6.2\times10^{5} \\ -3.1\times10^{4}}$ L$_{\odot}$ 
for the fitted-sizes case). We also look into differences between the average luminosity of clumps associated with MYSOs and UCH\,{\sc ii} regions and the latter turn out 
to be more luminous, with ${\overline{L_{IR, UCHII}}}$ = $9.6\times10^{5}\substack{+1.6\times10^{6} \\ -7.7\times10^{4}}$ L$_{\odot}$ 
(${\overline{L_{IR, UCHII}}}$ = $1.3\times10^{6}\substack{+1.9\times10^{6} \\ -8.2\times10^{4}}$ L$_{\odot}$ for the fitted-sizes case) and 
${\overline{L_{IR, MYSO}}}$ = $5.3\times10^{5}\substack{+7.8\times10^{5} \\ -4.3\times10^{4}}$ L$_{\odot}$ (${\overline{L_{IR, MYSO}}}$ = $6.7\times10^{5}\substack{+8.7\times10^{5} \\ -4.4\times10^{4}}$ L$_{\odot}$ 
for the fitted-size case), which is consistent with their more advanced evolutionary stage. This is also similar to what has been reported by Urquhart et al. (2014) 
who analyzed a sample of $\sim$ 800 ATLASGAL clumps associated with $\sim$ 1,000 RMS sources. These authors  compute their luminosities by using the model 
SED fitter developed by Robitaille et al. (2007) and by combining flux measurements from 2MASS (Skrutskie et al. 2006), UKIDSS (Lucas et al. 2008) 
or Vista-VVV (Minniti et al. 2010), MSX, WISE, Hi-Gal and ATLASGAL. They obtain that the luminosity distributions of the 
MYSO and UCH\,{\sc ii} regions subsamples are significantly different, with the two distributions peaking at $\sim$ 1$\times$10$^{4}$ L$_{\odot}$ 
and $\sim$ 4$\times$10$^{4}$ L$_{\odot}$,respectively.  This difference in the luminosity function was first discussed in Mottram et al. (2011b). 
Note that the lower luminosities reported by Urquhart et al. (2014) with respect to our values, are likely due to the fact that the ATLASGAL clumps are smaller
in size, i.e. on average $\sim$ 1.2 pc (see below for a comparison with our average size). 

\begin{figure*}
\centering
\includegraphics[width=12cm, height=14cm, angle=90]{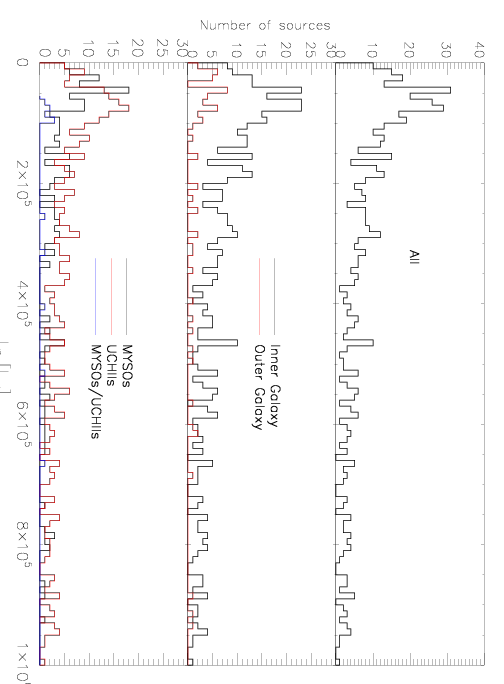}
\caption{Top panel: Overall distribution of luminosities for the complete sample. Middle panel: inner and outer Galaxy luminosity 
distribution. Bottom panel: MYSOs, UCH\,{\sc ii} regions and MYSOs/UCH\,{\sc ii} luminosity distribution.}
\end{figure*}

For the calculation of clump masses, assuming thermal equilibrium, we have:

\begin{equation}
M_{IR} = \frac{S{_{850\mu m}} D^2}{\kappa_{850\mu m} B_{\nu}(T_{d})}
\end{equation}

where the dust opacity, assuming a gas-to-dust ratio of 100, $\kappa_{850\mu m}$ = $\kappa_{353 \hspace*{0.1truecm} GHz}$ = 0.012 cm$^2$ g-1 (Preibisch et al. 1993), and T$_{d}$ = T$_{c}$.
Given that we only consider the cold dust component and ignore dust depletion effects in the source proximity, our computed mass will be an overestimate of the true values. 
Also, we emphasize that changing the value of the adopted dust opacity with affect the mass estimates. For instance, Ossenkopf $\&$ Henning (1994) give $\kappa_{700\mu m}$ = 0.0257 cm$^2$ g-1, 
causing masses to be a factor of two lower. Finally, the dust-to-gas ratio, rather than being a fixed value across the Galaxy, varies with Galactocentric distance and this will introduce 
further uncertanties in mass determinations.

Figure~6 illustrates the overall mass distribution (top panel), the inner and outer Galaxy mass distribution (middle panel), and the mass distribution for 
MYSOs, UCH\,{\sc ii} regions and MYSOs/UCH\,{\sc ii}. The resulting average mass is ${\overline{M_{IR}}}$ = $7.7\times10^{4}\substack{+1.3\times10^{4} \\ -0.9\times10^{3}}$ M$_{\odot}$ 
(${\overline{M_{IR}}}$ = $1.1\times10^{4}\substack{+1.8\times10^{4} \\ -1.0\times10^{3}}$ M$_{\odot}$  for the fitted-sizes case). The sources in the inner Galaxy are not only more luminous but also more 
massive compared to those in the outer Galaxy, in fact 
${\overline{M_{IR, inner}}}$= $8.2\times10^{3}\substack{+1.4\times10^{4} \\ -1.15\times10^{3}}$ M$_{\odot}$ (${\overline{M_{IR, inner}}}$ =  $1.2\times10^{4}\substack{+1.8\times10^{4} \\ -1.2\times10^{3}}$ M$_{\odot}$ 
for the fitted-sizes case) and 
${\overline{M_{IR, outer}}}$ = $2.2\times10^{3}\substack{+4.5\times10^{3} \\ -0.4\times10^{3}}$ M$_{\odot}$ (${\overline{M_{IR, outer}}}$ =  $3.3\times10^{3}\substack{+5.3\times10^{3} \\ -0.6\times10^{3}}$ M$_{\odot}$ 
for the fitted-sizes cases). Finally, clumps associated with UCH\,{\sc ii}  
regions are more massive than clumps associated to MYSOs, with ${\overline{M_{IR, UCHII}}}$ = $8.5\times10^{3}\substack{+1.4\times10^{4} \\ -1.1\times10^{3}}$ M$_{\odot}$ 
(${\overline{M_{IR, UCHII}}}$ = $1.2\times10^{4}\substack{+1.9\times10^{4} \\ -1.1\times10^{3}}$ M$_{\odot}$ for the fitted-sizes case) and ${\overline{M_{IR, MYSO}}}$ = $5.5\times10^{3}\substack{+9.9\times10^{3} \\ -0.7\times10^{3}}$ M$_{\odot}$ 
(${\overline{M_{IR, MYSO}}}$ = $6.8\times10^{3}\substack{+1.2\times10^{4} \\ -0.9\times10^{3}}$ M$_{\odot}$ for the fitted-sizes case), again in agreement with the former being older objects, 
which have undergone completely the accretion process. A similar result is reported in Urquhart et al. (2014) based on the analysis of the sample of ATLASGAL clumps (see above). 
They find that the MYSO and UCH\,{\sc ii} regions associated clump distributions have a median value of $\sim$ 1000 M$_{\odot}$ and 2800 M$_{\odot}$, respectively. Note, once again, that the lower ATLASGAL values 
can be ascribed to the smaller clump sizes. 

In the overall sample,  we have identified a sub-sample of 15 very massive objects, i.e. with a mass $>$ 4$\times$10$^{4}$ M$_{\odot}$. These are 3 MYSOs and 12 UCH\,{\sc ii} regions, which are  
concentrated towards {\em{l}} $\sim$ 43$^{\circ}$ and {\em{l}} $\sim$ 338$^{\circ}$, at a Galactocentric distance of, respectively, 7 kpc and 5 kpc. 

\begin{figure*}
\centering
\includegraphics[width=12cm, height=14cm, angle=90]{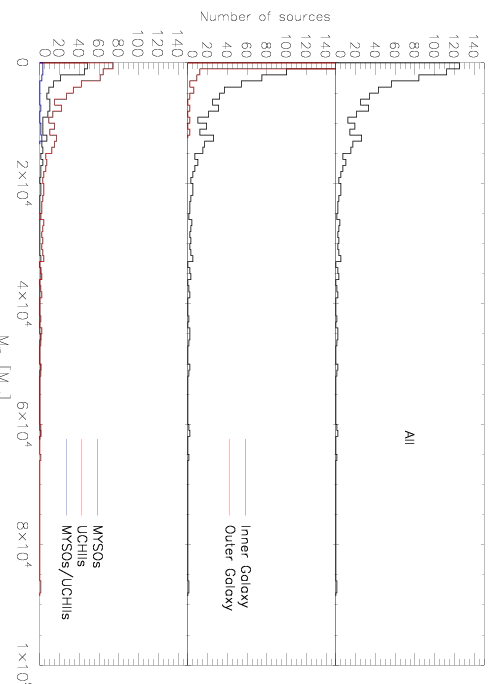}
\caption{Top panel: Overall distribution of mass for the complete sample. Middle panel: inner and outer Galaxy mass 
distribution. Bottom panel: MYSOs, UCH\,{\sc ii} regions and MYSOs/UCH\,{\sc ii} mass distribution.}
\end{figure*}

The derived masses allow us to estimate the sources surface densities, $\Sigma$ = M$_{IR}$ / ($\pi$ r$^{2}$), where $r$ is
the linear radius.  We obtain a mean surface density  equal to $0.25\substack{+0.17 \\ -0.003}$ g cm$^{-2}$ (Figure~7).
This mean value is greater than the characteristic surface densities of GMCs, $\Sigma \sim$ 0.035 g cm$^{-2}$ (Solomon et al. 1987), indicating that the clumps in our sample 
are gravitationally bound (Bertoldi $\&$ McKee 1992; Williams et al. 2000). At the same time this is lower than the average surface density (i.e. $\Sigma >$ = 1 g cm$^{-2}$) found by Plume et al. (1997)
for a sample of regions of massive star formation
observed in the $J$ = 5 $\rightarrow$ 4 and 2 $\rightarrow$ 1 transitions of CS and C$^{34}$S. Importantly, Plume's sources have a mean linear radius and (virial) mass of 0.5 pc and 
3800 M$_{\odot}$, respectively. Our sample is 
characterized by generally larger and more massive clumps, with a mean linear radius of 4.7$\pm$3.7 pc, and they do not represent the dusty counterparts of the
molecular dense clumps studied by Plume et al. (1997): while Plume et al.'s observations trace the densest material directly involved in the production of massive stars,
the {\em{Planck}} measurements are sensitive to the total mass involved, including less dense material. More recent studies 
(Butler $\&$ Tan 2012; Traficante et al. 2018, 2020)
have revised down the massive star formation threshold value and today it is generally accepted that this is to be found in the range 0.1 - 0.35 g cm$^{-2}$, in agreement with the fact 
that our clumps are forming massive stars.

\begin{figure}[h]
\centering
\hspace*{-0.5truecm}
\includegraphics[width=9cm, height=7cm, angle=0]{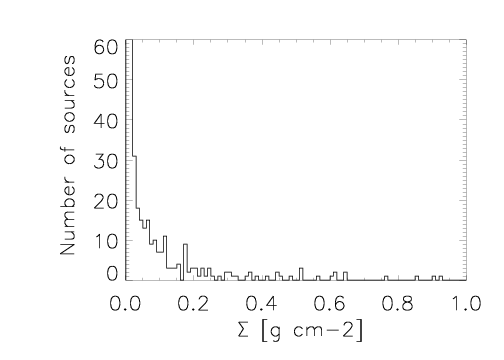}
\caption{Surface density distribution for the clumps in the complete sample.}
\end{figure}

\section{Variation of the Star Formation Efficiency with Galactocentric Radius}

It is well established that the outer Galaxy presents a very different environment with respect to the inner
Galaxy, being the former characterized by low total gas densities, low metallicities and gas abundances, as well as by higher shear. Because of the Kennicutt (1998) law,   
the low densities makes the gas gravitationally stable, hence not apt to star formation. Despite these unfavorable environmental conditions, sporadic star formation is  observed
towards the outer Galaxy, suggesting that other mechanisms, other than gravity, may trigger the star formation process. In particular, Elmegreen $\&$ Hunter (2006)
propose that some degree of turbulence persists in the outer disk, allowing the formation of clouds and compensating for the lack of gravitational instabilities. 

A measure of the global {\em{instantaneous}} SFE can be obtained from the ratio of the IR luminosity, $L_\mathrm{IR}$, to
clump mass, $M_\mathrm{IR}$. Both these quantities have the same square dependence on distance (D$^{2}$), therefore their ratio is
distance independent. This is very important given that solar distances, especially when derived from kinematic measurements, can be affected by large uncertainties.
Using the same binning in Galactocentric radii described in Section~4.1, we then investigate how the luminosity-to-mass ratio, $L_\mathrm{IR}$/$M_\mathrm{IR}$, varies from
the inner to the outer Galaxy. For this purpose, we first analyze the $L_\mathrm{IR}$/$M_\mathrm{IR}$ distributions per Galactocentric bin (Figure~8).
Then, we average the $L_\mathrm{IR}$/$M_\mathrm{IR}$ values in each bin, and look at the distribution of the mean values as a function of Galactocentric radius (Figure~9 and Table~3). 
It is apparent from Figure~8 that there is a significant dispersion of the  $L_\mathrm{IR}$ to $M_\mathrm{IR}$ ratios in each bin, indicating that star 
formation does not simply scale with distance from the Galactic center, but instead
is a complex process which depends on many parameters. If the determing factors in defining the star formation activity at a given location in the Galaxy were quantities which linearly scale 
with the Galactocentric radius, such as gas column density and metallicity, we would expect a much tighter correlation. Despite the large scatter in every bin, a global trend emerges from 
both Figure~8 and Figure~9: the mean luminosity-to-mass ratio decreases towards the outer Galaxy, 
with a difference between the most three inner bins (2 kpc $<$ R$_G$ $<$  8 kpc) and the most three outer ones (8 kpc $<$ R$_G$ $<$ 14 kpc) of the order of 53$\%$.

\begin{figure*}
\centering
\includegraphics[width=14cm, height=16cm, angle=90]{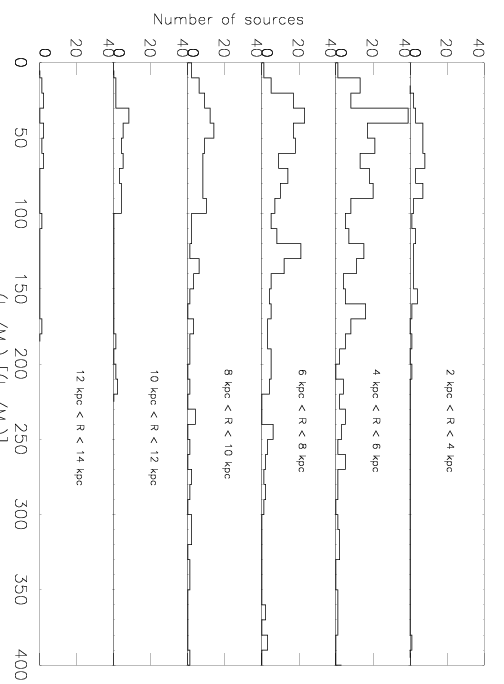}  
\vspace*{0.3truecm}
\caption{IR luminosity to dust mass ratio ($L_\mathrm{IR}$/$M_\mathrm{IR}$) as a function of Galactocentric radius, R$_{G}$.} 
\end{figure*}

Similarly, if one considers the $L_\mathrm{IR}$/$M_\mathrm{IR}$ distributions for the {\em{inner}}/{\em{outer Galaxy}} sources, using the same convention for defining
the two sub-samples described in Section~4.1, we obtain: ${\overline{(L_{IR,i}/M_{IR,i})}}$ = $119.1\substack{+177.6 \\ -34.2}$ L$_{\odot}$/M$_{\odot}$ (${\overline{(L_{IR,i}/M_{IR,i})}}$ = 
$118.9\substack{+169.4 \\ -31.9}$ L$_{\odot}$/M$_{\odot}$ for the fitted-sizes case) and
${\overline{(L_{IR,o}/M_{IR,o})}}$ = $109.8\substack{+1150.8 \\ -43.1}$  L$_{\odot}$/M$_{\odot}$ (${\overline{(L_{IR,o}/M_{IR,o})}}$ = $94.9\substack{+135.3 \\ -33.7}$ L$_{\odot}$/M$_{\odot}$ for the fitted-sizes case), 
where $L_\mathrm{IR,i}$/$M_\mathrm{IR,i}$ and $L_\mathrm{IR,o}$/$M_\mathrm{IR,o}$ are the IR luminosity to clump mass ratio for the {\em{inner}} and {\em{outer Galaxy}} samples, respectively (Figure~10).

\begin{figure*}
\centering
\epsfig{file=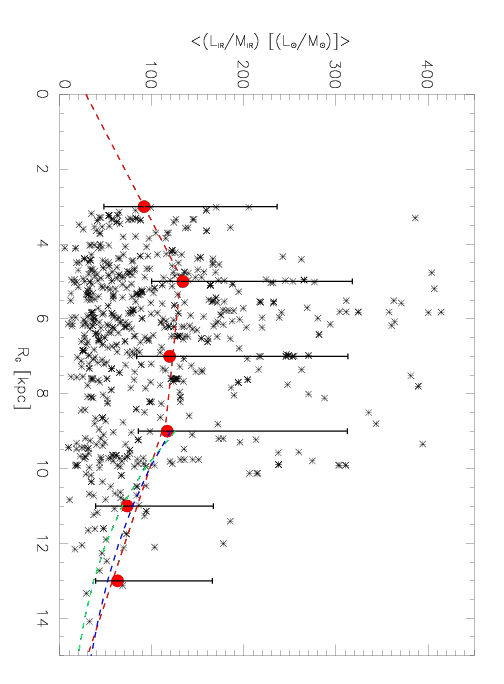, width=11cm, height=15cm, angle=90}
\caption{IR luminosity ($L_\mathrm{IR}$) to dust mass ($M_\mathrm{IR}$) ratio, $L_\mathrm{IR}$/$M_\mathrm{IR}$, as a function of Galactocentric radius, R$_{G}$. The error bars
are the 15.9$\%$ and 84.1$\%$ percentiles. The best-fit to the data is shown (red-dashed line). Also shown is the exponential decay from Leroy et al. for the case of an optical 
radius of 13 kpc (green-dashed line) and 19 kpc (blue-dashed line). The stars represent the individual data points.}
\end{figure*}

If we look closely at Figure~8 and Figure~9, we notice that the SFE, instead of uniformly decreasing with R$_{G}$, first increases at small radii (R$_G$ $<$ 4 kpc) and then peaks at R$_G$ $\sim$ 5 kpc, 
after which it linearly decreases. This behavior can overall be explained by considering the 
molecular fraction radial profile as discussed, for instance, in Koda, Scoville $\&$ Heyer (2016). These authors show that, from a quantitative point of view, the fraction of molecular gas 
decreases from 100$\%$ near the center of the Galaxy, dropping to 50$\%$ at 6 kpc, and decreasing to 10-20$\%$ at $R \sim$ 8.5 kpc. The prominent bump at $\sim$ 4.5 kpc is due to 
the bar structure of our Galaxy, with a half-length of 4.4$\pm$0.5 kpc (Benjamin et al. 2005), and this is associated to bright CO at the end of the bar, 
as observed in many external barred spiral systems (e.g. Sheth et al. 2002). 

The correlation between SFE and molecular gas is a longstanding problem discussed in the literature, and even more so is the role played 
by spiral arms. Theoretical models (e.g. Dobbs et al. 2006) predict that spiral arms play a pivotal role in the formation of GMCs and 
enhance the SFE. However, so far we were lacking evidence of this. Previous efforts carried out by Moore  et  al. (2012) in the first Galactic quadrant found 
that the $L_\mathrm{IR}$/$M_\mathrm{IR}$ ratio was relatively flat for the inner 5 kpc. Two significant peaks were detected at $\sim$ 6 kpc and $\sim$ 8 kpc, respectively, 
but these were attributed to the presence of of the star formation complexes W51 snd W49. The work of Moore et al. (2012) was later extended by Eden et al. (2013; 2015) who reported no 
significant variation of the SFE between different arms or the inter-arm regions. More recently, Urquhart et al. (2018, 2020), in analyzing the ATLASGAL sample in the first and 
fourth quadrant, found  that the SFE as described by the L$_{bol}$/M$_{clump}$ ratio is relatively flat between 2 and 9 kpc when evaluated over kiloparsec scales, although local 
enhancements can be detected on smaller scales, 
in agreement with Moore et al. (2012) and Eden et al. (2013; 2015). This body of work led all these authors to conclude that: "{\em{the spiral arms are principally collecting material 
together via orbit crowding but there is no evidence that they are playing a role in enhancing the star formation within molecular clouds.}}"

We believe that we are detecting correlations of the SFE with Galactocentric distance, and with the overall distribution of molecular material, because, for the 
first time, we are combining three ingredients: 1) a complete sample of objects; 2) the exploration of the Galactic plane as a whole; 3)  the 
use of data not affected by spatial filtering.

Figure~9 also shows that the peak at R$_G$ $\sim$ 5 kpc is not followed by a steady decrease in SFE. Instead, 
the IR luminosity, $L_\mathrm{IR}$, to clump mass, $M_\mathrm{IR}$, ratio remains almost constant up to $\sim$ 9 kpc. 
A best-fit to $L_\mathrm{IR}$/$M_\mathrm{IR}$ as a function of Galactocentric radius gives:\\

\[
    L_{IR}/M_{IR}=\left\{       \tag{4}
                \begin{array}{ll}
                  28.3 + 21.1 \times R_{G}   \hfill  R_{G} \hspace*{0.1truecm} < \hspace*{0.1truecm} 5 \hspace*{0.1truecm} \mathrm{kpc}\\
                  152.8 - 4.2 \times R_{G}  \hspace*{0.3truecm}  \hfill  5 \hspace*{0.1truecm} \mathrm{kpc} \hspace*{0.1truecm} < \hspace*{0.1truecm} R_{G} \hspace*{0.1truecm} < \hspace*{0.1truecm} 9 \hspace*{0.1truecm} \mathrm{kpc}\\
                  232.6 - 13.5 \times R_{G}  \hspace*{0.2truecm}  \hfill R_{G} \hspace*{0.1truecm} > \hspace*{0.1truecm} 9 \hspace*{0.1truecm} \mathrm{kpc}
                \end{array}
              \right.
  \]

\vspace*{0.2truecm}

Importantly, the best-fit solution is obtained by folding in the spread in the data points in each bin. We notice that the two last bins contains only few data points, i.e.
39 and 11, respectively. However, since the sample is complete, these bins are significant and their weight in the fit is the same as the other bins.

Leroy et al. (2008), based on a study of 23 external spiral and dwarf galaxies, find that in spiral galaxies the SFE appears approximately constant up to the optical radius (r$_{25}$), after
which it declines exponentially. Taking into account that for our Galaxy the optical radius is estimated to be at $\sim$ 13 - 19 kpc (Ruffle et al 2007; Sale et al. 2010), the Leroy 
et al.'s relation for the Milky Way can be re-written as:

\[
   SFE \propto L_{IR}/M_{IR}=\left\{   \tag{5}
                \begin{array}{ll}
                  const   \hfill  R_{G} \hspace*{0.1truecm} < \hspace*{0.1truecm} 5.6 \hspace*{0.1truecm} - \hspace*{0.1truecm} 8.2 \hspace*{0.1truecm} \mathrm{kpc}\\
                  \sim \hspace*{0.1truecm} e^{-R_{G}/(3.25 \hspace*{0.1truecm} - \hspace*{0.1truecm} 4.75 \hspace*{0.1truecm} \mathrm{kpc})} \hfill R_{G} \hspace*{0.1truecm} > \hspace*{0.1truecm} 5.6 \hspace*{0.1truecm} - \hspace*{0.1truecm}$ 8.2 \hspace*{0.1truecm}$ \mathrm{kpc}  
                \end{array}
              \right.
  \]

\vspace*{0.2truecm}

Figure~9 illustrates the linear fit to the data (equation~4) and the Leroy et al.'s exponential solutions for two values of the optical radius (equation~5). 
Leroy et al. point out that the ISM is equal parts of HI and H2 at R$_{G}$ = 0.4 $\times$ r$_{25}$ $\pm$ 0.18. Therefore their result 
can be interpreted as a steady decrease of the SFE beyond the HI-to-H2 transition radius. However, as we see from Figure~9, in the case of our Galaxy the SFE remains fairly constant 
up to R$_{G}$ $\sim$ 9- 10 kpc, suggesting either the presence of a significant, previously undetected, molecular material at R$_G$ $>$ 8 kpc and/or 
that the radial dependence of the SFE may not be simply a function of the HI-to-H2 transition radius as previously thought.

\begin{figure}
\centering
\epsfig{file=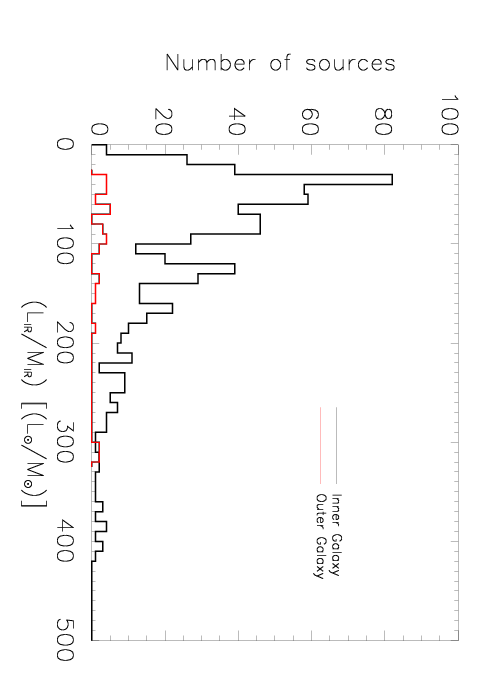, width=7cm, height=9cm, angle=90}
\caption{Luminosity-to-mass ratio in the inner (first and fourth) and outer (second and third) Galactic quadrants.
}
\end{figure}

From the observational point of view, the former finds at least partial confirmation in the hydrogen recombinations lines (H86$\alpha$ to H93$\alpha$)
and radio continuum (9 GHz) survey carried out by Anderson $\&$ Bania (2011) with the Green Bank Telescope (GBT). These observations have allowed the discovery, in the first Galactic quadrant, 
of a new population of 34 HII regions located in the Outer Arm  (Bania et al. 2010), at $R_{G} >$ 8 - 9 kpc. Previously, there were only 7 HII regions known in this area. 


Finally, we have investigated variations of the SFE as a function of Galactocentric radius for the two populations of MYSOs and UCH\,{\sc ii} regions, separately. 
A behavior similar to the parent population is found for each of these evolutionary stages (see Table~3). 

\begin{table*}
\begin{center}
\caption{Average $L_\mathrm{IR}$/$M_\mathrm{IR}$ per Galactocentric bin.}
\begin{tabular}{cccccc}
\hline
\hline
 bin center    & n. sources   & Log($L_\mathrm{IR}$/$M_\mathrm{IR}$) (unresolved) & Log($L_\mathrm{IR}$/$M_\mathrm{IR}$) (fitted sizes) &   Log($L_\mathrm{IR}$/$M_\mathrm{IR}$) (UCH\,{\sc ii}) &     Log($L_\mathrm{IR}$/$M_\mathrm{IR}$) () \\
  (kpc)   &              &    Log(L$_{\odot}$/M$_{\odot}$)   &   Log(L$_{\odot}$/M$_{\odot}$) &  Log(L$_{\odot}$/M$_{\odot}$) & Log(L$_{\odot}$/M$_{\odot}$) \\
          &              &                                   &             &           &    \\
   3  &     57       &  $91.5\substack{+144.5 \\ -43.7}$   &  $83.2\substack{+126.2 \\ -44.0}$  & $81.4\substack{+176.8 \\ -39.3}$ & $83.6\substack{+126.2 \\ -44.6}$\\
   5  &   274        &  $133.6\substack{+184.2 \\ -33.9}$  & $141.8\substack{+183.5 \\ -33.1}$ &  $104.6\substack{+163.2 \\ -28.6}$ & $157.9\substack{+203.5 \\ -33.2}$\\
   7  &   224        &  $119.2\substack{+193.8 \\ -35.8}$  &  $115.6\substack{+174.1 \\ -31.9}$ & $93.2\substack{+152.7 \\ -28.9}$ & $124.1\substack{+180.9 \\ -34.2}$ \\
   9  &    126        &  $116.7\substack{+195.6 \\ -31.3}$ &  $107.3\substack{+208.9 \\ -30.7}$ & $88.5\substack{+171.6 \\ -23.2}$ & $113.9\substack{+223.9 \\ -31.9}$\\
  11 &    39    &     $73.1\substack{+93.6 \\ -34.2}$ &    $65.5\substack{+84.8 \\ -30.7}$ & $52.3\substack{+77.7 \\ -28.3}$ &  $72.9\substack{+162.3 \\ -33.7}$\\
  $>$13 &    11    &   $62.7\substack{+102.9 \\ -23.9}$ &   $57.5\substack{+107.2 \\ -23.9}$ & $62.8\substack{+145.8 \\ -16.3}$ &  $52.2\substack{+107.2 \\ -23.9}$\\
\hline
\hline
\end{tabular}
\end{center}
\end{table*}

\section{Conclusions}

{\em{Planck}} whole-sky sub-millimeter maps provide an unprecedented opportunity to carry out an unbiased study
of the environmental conditions in which massive star formation takes place. Using {\em{Planck}} HFI upper frequency bands
(350, 500, 850 $\mu$m) complemented by IRAS/IRIS 100, 60 and 25$\mu$m bands, we have estimated dust temperatures, luminosities, masses and
surface densities of clumps associated to a complete sample of candidate MYSOs and UCH\,{\sc ii} regions selected from the RMS survey data base.

Exploting the {\em{Planck}} and IRAS/IRIS full-coverage of the Galactic Plane, we have searched for variations of the clump dust temperatures (warm and cold component)
with respect to distance from the Galactic center. We find that the distribution
of the cold dust temperature component of the clumps generally reflects the large scale variation of dust temperature for the ISM along the Galactic Plane, that is: massive star complexes
at smaller Galactocentric radii have higher temperatures than complexes at larger radii. An opposite trend is found for the warm dust component that appears to increase with distance 
from the center of the Galaxy, suggesting that the more embedded dust is subjet to the local conditions rather than to the ISRF.   

We have also explored how the luminosity-to-mass ratio (L/M) associated with the clumps varies from the inner to the outer Galaxy. Our results are consistent 
with a SFE decreasing from the inner to the outer Galaxy. However, the cut-off radius that identifies the transition between a fairly constant and a declining SFE is at a 
larger radius (R$_G$ $\sim$ 9 kpc) than expected. This may have implications for the amount of molecular gas still undetected 
at R$_G$ $>$ 8 kpc. 

Finally, we present in the Appendix color-color and color-magnitude plots of MYSOs and UCH\,{\sc ii} regions compared to cold clumps from the PGCC. These colors 
will be useful for identification of sites of massive star formation in future large-scale surveys.

\begin{acknowledgements}

This paper is based and made use of information from the Red MSX
Source survey database at www.ast.leeds.ac.uk/RMS which was
constructed with support from the Science and Technology Facilities
Council of the UK.

\end{acknowledgements}

\appendix

The photometric measurements described in Section~4, as well as allowing us to derive the physical properties of the sources presented above, also make it
possible to generate color-color and color-magnitude plots, the purpose of which is to serve as diagnostic tools for the identification of MYSOs and UCH\,{\sc ii} regions in blind
Galactic surveys. To this end, we have performed a random selection from the {\em{Planck}} Catalog of Galactic Cold Clumps (PGCC, Planck Collaboration XXVIII 2016) and compared the
colors/magnitudes of the randomly extracted cold clumps to those of the complete sample of MYSOs and UCH\,{\sc ii} regions from the RMS database.
The PGCC contains 13,188 Galactic cold sources (median temperature between 13 and 14.5 K) spread across the whole sky.

Figure~11 shows that the MYSOs and UCH\,{\sc ii} regions (red triangles) and cold clumps (blue triangles) occupy very distint regions in the color-color and
color-magnitude plots. The  MYSOs and UCH\,{\sc ii} regions are found for: 

\begin{center}

\begin{equation}
-0.5 < Log(F_{100}/F_{350}) < 4., -2.5 < Log(F_{250}/F_{850}) < 0. 
\end{equation}

\end{center}

The two populations are characterized by different dust temperatures that cause this segregation: as we have discussed in
Section~4.1, there are two temperature dust components to the the MYSOs and UCH\,{\sc ii} regions SED, one warm, of the order of 30 - 50 K, and one 
cold, around 20 K. On the contrary, cold clumps present only one dust temperature component, i.e. the cold one. This implies that the SED of MYSOs and UCH\,{\sc ii} regions declines rapidly between
250 and 850 $\mu$m and between 100 and 350 $\mu$m, while in these wavelength ranges, the spectrum of a cold clump decreases less rapidly (250 $< \lambda <$ 750 $\mu$m) or even still rises
(100 $< \lambda <$ 350 $\mu$m).

\begin{figure*}{H}
\centering
\includegraphics[width=6.5cm, height=8cm, angle=90]{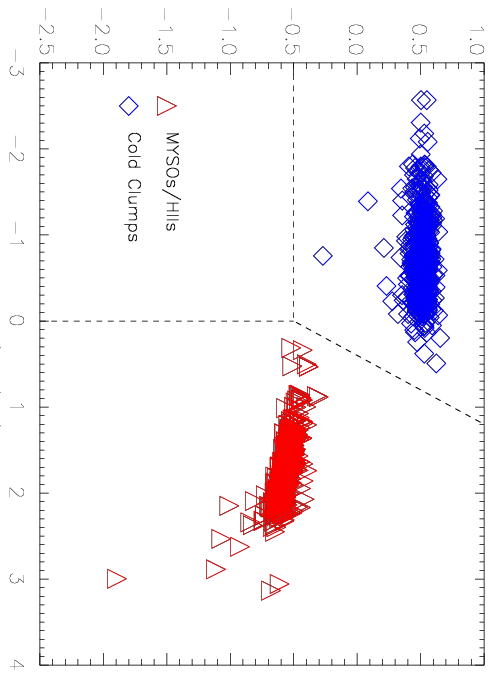}
\includegraphics[width=6.5cm, height=8cm, angle=90]{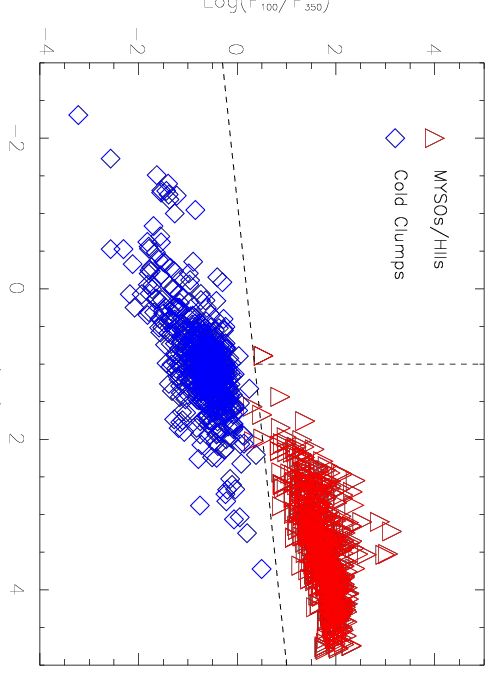}
\vspace*{0.3truecm}
\caption{Color-color (left) and color-magnitude (right) plots for MYSOs and UCH\,{\sc ii} regions from the RMS database (red triangles) and
cold clumps from the PGCC (blue triangles).}
\end{figure*}

\end{document}